\documentclass[preprint,amsmath,amssymb,showkeys,showpacs]{revtex4}
\usepackage{graphicx}
\usepackage{epsfig}
\usepackage{amssymb}

\begin{document}

\title{Short-time dynamics of finite-size mean-field systems}

\author{Celia Anteneodo}
%\email{celia@fis.puc-rio.br}
\address{Departmento de F\'{\i}sica, PUC-Rio and
National Institute of Science and Technology for Complex Systems,
Rua Marqu\^es de S\~ao Vicente 225, G\'avea, CEP 22453-900 RJ,
Rio de Janeiro, Brazil}

\author{Ezequiel E. Ferrero}
%\email{ferrero@famaf.unc.edu.ar}
\address{Facultad de Matem\'atica, Astronom\'{\i}a y F\'{\i}sica,
Universidad Nacional de C\'ordoba and   Instituto de F\'{\i}sica Enrique
Gaviola  (IFEG-CONICET),
Ciudad Universitaria, 5000 C\'ordoba, Argentina}

\author{Sergio A. Cannas}
%\email{cannas@famaf.unc.edu.ar}
\address{Facultad de Matem\'atica, Astronom\'{\i}a y F\'{\i}sica,
Universidad Nacional de C\'ordoba and   Instituto de F\'{\i}sica Enrique
Gaviola  (IFEG-CONICET),
Ciudad Universitaria, 5000 C\'ordoba, Argentina}

\begin{abstract}
We study the short-time dynamics of a mean-field model with non-conserved order
parameter (Curie-Weiss with Glauber dynamics) by solving the associated Fokker-Planck equation.
We obtain closed-form  expressions for the first moments of the order parameter,
near to both the critical and spinodal points, starting from different initial conditions. This allows us to confirm the
validity of the short-time dynamical scaling hypothesis in both cases.
Although the procedure is illustrated for a particular mean-field model,
our results can be straightforwardly extended to generic models with a single
order parameter.
\end{abstract}

 

\maketitle

%%%%%%%%%%%%%%%%%%%%%%%%%%%%%%%%%%%%%%%
\section{Introduction}

Universal scaling behavior appears to be an ubiquitous property of critical
dynamic systems.  While initially believed to hold only in the long time limit,
it was realized during the last decade that the dynamical scaling hypothesis
can be extended to the short-time limit~\cite{JaScSc1989}.
This is accomplished by assuming that, close to the critical point, the $nth$
moment of the order parameter obeys the homogeneity relation
\begin{equation} \label{scaling1}
    m^{(n)}(t,\tau,L,m_0) = b^{-n\beta/\nu} m^{(n)}(b^{-z}t,b^{1/\nu}\tau,L/b,b^\mu m_0),
\end{equation}
where $t$ is time, $\tau$ is the reduced temperature
$\tau=(T_c-T)/T_c$, $L$ is the linear system size, $m_0$ is the initial value
of the order parameter  and $b$ is a spatial
rescaling parameter. $\mu$ is a universal exponent that describes
the short-time behavior, while $\beta$, $\nu$, and $z$ are the usual
critical exponents. When $m_0<<1$ we recover the usual dynamic scaling relation,
from which a power law relaxation at the critical point (for instance,
in the magnetization $n=1$) $m(t) \sim t^{-\beta/\nu z}$ when  $L\gg1$
and $t \gg 1$  follows. This is the critical slowing down.

On the other hand, the short-time dynamics (STD) scaling properties of the system
depend on the initial preparation,  i.e., on the scaling field $m_0$.
Setting $b=t^{1/z}$, from Eq.~(\ref{scaling1}) one obtains for small (but non-null)
values of $t^{\mu/z}m_0$ in the large $L$ limit
\begin{equation}
  \label{eq:short-time-m-scaling}
  m(t,\tau,m_0) \sim m_0\, t^\theta F(t^{1/\nu z}\tau), \qquad \theta =
  \frac{\mu-\beta/\nu} {z}.
\end{equation}
Hence, at the critical point $\tau=0$ an initial increase of
the magnetization $m\sim m_0 t^\theta$ is observed. For the second moment $n=2$
 the dependency on $m_0$ can be neglected when $m_0 \ll 1$.
Since $m^{(2)}\sim N^{-1}=L^{-d}$ in the large $L$ limit ($d$ is the spatial
dimension), one obtains at the critical point
\begin{equation}
  \label{eq:scaling-second-moment}
  m^{(2)}(t) \sim N^{-1}\, t^{d/z-2\beta/z\nu}.
\end{equation}

The short-time universal scaling behavior  has been verified  in a large variety
of critical systems both by renormalization group (RG) calculations~\cite{JaScSc1989,PrPrKaTs2008} and
Monte Carlo (MC) numerical simulations~\cite{Zh1998,Zh2006,PrPrKrVaPoRy2010}.
The hypothesis also applies when the system starts in the completely ordered
state, i.e., $m_0=1$. In this case it is assumed that the homogeneity relation
\begin{equation}\label{scaling2}
    m^{(n)}(t,\tau,L) = b^{-n\beta/\nu} m^{(n)}(b^{-z}t,b^{1/\nu}\tau,L/b)
\end{equation}
holds even for short (macroscopic) time scales. Hence, in the large $L$ limit we have
\begin{equation}
  m(t) = t^{-\beta/\nu z} G(t^{1/\nu z}\tau),
  \label{eq:short-time-m-ordered}
\end{equation}
and taking the derivative of $\log m$,
\begin{equation}
  \left.\frac{\partial \log m(t,\tau)}{\partial \tau}\right|_{\tau=0}
  \sim t^{1/\nu z}.
  \label{eq:scaling-log-m}
\end{equation}

 While the scaling hypothesis starting from the disordered state is supported
 both by numerical simulations and RG, its validity for an initial ordered
 state relies up to now only on numerical simulations.

Recently, numerical simulations have shown that the short-time scaling
hypothesis (\ref{scaling1}) holds not only close to a critical point,
but also close to spinodal points in systems exhibiting a first-order
phase transition, both for mean-field and short-range interactions
models~\cite{LoFeCaGr2009}. This is particularly interesting, because
it suggests the existence of some kind of diverging correlation length
associated to a spinodal point. Since the proper concept of spinodal
in short-range interactions  systems is still a matter of debate
(see Ref.~\cite{LoFeCaGr2009} and references therein),  a deeper
understanding of the microscopic mechanisms behind the observed short
time scaling could shed some light on this problem. One way of
achieving this goal is to look for exact solutions of particular  models.
A first step in that direction is to analyze mean-field (i.e.,
infinite-range interactions) models, for which the concept of spinodal is well
defined~\cite{LoFeCaGr2009}. That is the objective of the present work:
we analyze the exact STD behavior of far from equilibrium
mean-field systems with non-conserved order parameter.

Non-equilibrium phenomena in physics and other fields are commonly
studied through Fokker-Planck equations (FPEs).
In particular, non-equilibrium dynamical aspects of phase transitions
can be analyzed by means of the FPE associated to the
master equation describing the microscopic dynamics~\cite{binder,hanggi,munoz}.
In fact, this tool was proved to be useful in the description of
the relaxation of metastable states~\cite{binder},
finite-size effects~\cite{ruffo} or the impact of fluctuations in control
parameters~\cite{politi}, and have been considered for mean-field
spin models~\cite{binder,mori} and
coupled oscillators~\cite{ruffo},   amongst many others.

As soon as the degrees of freedom of the system can be reduced
to a few relevant ones, a  low-dimensional FPE can be found.
Although this description is suitable for properties that do not depend
on the details of the dynamics,
or  for mean-field kinetics, many conclusions are expected to hold in more
general instances.

For a single order parameter $m$, the FPE for its probability $P=P(m,t|m_0,0)$ is
\begin{equation} \label{FPE}
\partial_t P =\left[ -\partial_m D_1(m) +\partial_{mm} D_2(m) \right]
P \equiv L_{FP}(m)P  \,,
\end{equation}
where the drift and diffusion coefficients are determined by the
Hamiltonian and the particular dynamics (e.g., Glauber or Metropolis).

Following this stochastic approach, here we study the scaling of the short-time
relaxational dynamics in the vicinity of critical and spinodal points.
In first approximation,  the drift $D_1(m)$ ($=-dV/dm$) is generically linear
in the vicinity of a critical point and quadratic in the spinodal,
following the quadratic and cubic behavior of the drift potential $V$,
respectively.
Meanwhile, typically in various models, the noise intensity $D_2(m)$ scales as
$\epsilon\sim1/N$~\cite{binder,ruffo}.
Therefore, although we will present the  STD  for a
particular spin model, our results
can be straightforwardly extended to more general mean-field ones.

%%%%%%%%%%%%%%%%%%%%%%%%%%%%%%%%%%%%%%%
\section{Formal FPE solution and moment expansions}

The formal solution of the FPE~(\ref{FPE}), for the initial condition
$P(m,0|m_0,0)=\delta(m-m_0)$,
is~\cite{risken}
$$  
P(m,t|m_0,0) ={\rm e}^{t\,L_{FP}(m)}\delta(m-m_0) \,.
$$

The average of an arbitrary quantity $Q(m)$
can be derived directly from the FPE,
by means of the adjoint
Fokker-Plank operator $L^\dag_{FP}(m) \equiv D_1 \partial_m  +D_2\partial_{mm} $,
as follows
\begin{eqnarray}  \nonumber
\langle Q\rangle(m_0,t) &=&\int Q(m)\,P(m,t|m_0,0) dm   
=\int Q(m){\rm e}^{t\,L_{FP}(m)}\delta(m-m_0)  dm    \\ \nonumber
 &=&\int  \delta(m-m_0) {\rm e}^{t\,L^\dag_{FP}(m)} Q(m) dm    
=  {\rm e}^{t\,L^\dag_{FP}(m_0)} Q(m_0) = \\ \label{adjoint}
&=&\sum_{k\ge 0} [L^\dag_{FP}(m_0)]^k Q(m_0)\,t^k/k!  \,.
\end{eqnarray}
Therefore, the first two moments of the order parameter are
\begin{eqnarray} \nonumber
\langle m \rangle &=&  m_0 + D_1 t + \frac{1}{2}[ D_1D_1^\prime+D_2D_1^{\prime\prime}] t^2 +\ldots\,,
\\ \label{M1M21storder}
\langle  m^2 \rangle &=&
 \langle m \rangle^2  +  2D_2t+[ 2D_2D_1^\prime +D_1D_2^\prime+D_2D_2^{\prime\prime}  ]t^2 +\ldots \,,
\end{eqnarray}
where $D_1,\; D_2$ and their derivatives are evaluated in $m_0$.
Notice that if $D_1$ and $D_2$ are not state-dependent, the expansion up to
first order is exact.

Alternatively, evolution equations for moments can be obtained by integration of
Eq.~(\ref{FPE}), after multiplying each member of the equation by the quantity to be averaged, that is

\begin{equation}\label{moments}
    \frac{d \langle m^n  \rangle }{dt}=  n\,\langle  m^{n-1} D_1(m) \rangle + n(n-1)\,\langle m^{n-2} D_2(m) \rangle \,.
\end{equation}

\noindent For $n=1$ we have

\begin{equation}\label{eqM1}
    \frac{d \langle m  \rangle }{dt}=  \langle  D_1(m) \rangle \,.
\end{equation}

Eqs.(\ref{moments}) lead in general to a hierarchy of coupled equations  for the moments. 
Only for a few special cases ($D_1$ and $D_2$ polynomials in $m$ of degree smaller or 
equal than one and two respectively) these equations decouple. Otherwise, one has to rely  
on approximated methods to solve their dynamics.
%

%%%%%%%%%%%%%%%%%%%%%%%%%%%%%%%%%%%%%%%
\section{Paradigmatic mean-field model}

Let us exhibit our STD analysis for the paradigmatic system of $N$
fully connected Ising spins (Curie-Weiss model), subject to a magnetic field $H$, ruled
by the mean-field Hamiltonian
\begin{equation}
{\cal H}= -\frac{J}{2N}M^2-HM.
\end{equation}

Since the Hamiltonian depends only on the total magnetization $M$, the master equation for this
model can be written in closed form for $M$~\cite{binder,mori}.
In the large $N$ limit, when the magnetization per spin $m=M/N$ can be taken
as a continuous variable, an expansion  of the master equation up to first order in
the perturbative parameter $\epsilon=1/N$ leads for the Glauber dynamics  to a
FP equation (\ref{FPE}) with~\cite{mori}
\begin{eqnarray} \nonumber
D_1(m) &=& -m+\tanh[m^\prime]-\epsilon \beta J m \,{\rm sech}^2[m^\prime] \,,\\
\label{D1D2}
D_2(m) &=& \epsilon\bigl(  1- m\tanh[m^\prime]  \bigr) \,,
\end{eqnarray}
where we have defined $m^\prime=\beta(Jm+H)$, with $\beta=1/(k_B T)$.

In the next sections we derive asymptotic solutions of the FPE with these coefficients,
both close to the critical point ($H=0$ and $T \approx T_c=J/k_B$) and to spinodal points
for $T< T_c$. Analytical results are compared against Monte Carlo simulation ones using
Glauber algorithm.
Time was adimensionalized with  the characteristic time $t_0$ of the
transition rate $w= t_0^{-1}(1+\exp(\beta\Delta{\cal H}))^{-1}$. The unit of time
in theoretical expressions corresponds to one MC step in simulations.
We also performed several checks using Metropolis algorithm. The
outcomes were indistinguishable from the Glauber ones,
except for a trivial time rescaling factor 2 close to the critical point,
as expected~\cite{binder}.

%%%%%%%%%%%%%%%%%%%%%%%%%%%%%%%%%%%%%%%
\section{STD near the critical point}

In the vicinity of the critical point (at $T\simeq T_c=J/k_B\equiv 1$ and $H=0$),
the coefficients (\ref{D1D2})  can be approximated for small $m$
(i.e., $\beta J |m|<<1$) respectively by
\begin{eqnarray} \nonumber
D_1(m) &=& -\omega(\lambda,\epsilon)\,m-\kappa(\lambda,\epsilon)m^3
+ {\cal O}(m^5) \,,\\ \label{D1D2_crit}
D_2(m) &=& \epsilon\left( [1-(1-\lambda) m^2] +{\cal O}(m^4)\right) \,,
\end{eqnarray}
where $\omega(\lambda,\epsilon) \equiv  \lambda + \epsilon (1-\lambda) $ and
$\kappa(\lambda,\epsilon) \equiv (\frac{1}{3}-\epsilon)(1-\lambda)^3$,
with  $\lambda\equiv 1-T_c/T$.

\begin{figure}[b]
\centering
\includegraphics[bb=60 360 540 720, width=0.6\textwidth]{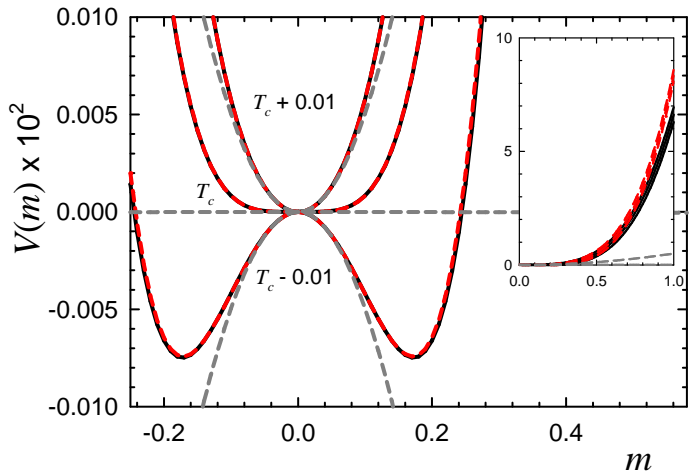}
\caption{(Color online) Potential $V(m)$, for different values of $T$ in
the vicinity of $T_c$, indicated in the figure.
It was obtained (unless an arbitrary additive constant) from
the  integration of $D_1$ in Eqs.~(\ref{D1D2}) (black full lines),   (\ref{D1D2_crit})
(red dashed lines) and (\ref{D1linear}) (gray dashed lines).
Inset: zoom of the region close to $m=1$.
}
\label{fig:pot1}
\end{figure}

Within the domain of validity of these approximations
$(1-\lambda)m^2<<1$ and therefore $ D_2\simeq \epsilon.$
Concerning $D_1$, its linear term dominates, that is,
\begin{equation} \label{D1linear}
D_1(m)\simeq  -\omega(\lambda,\epsilon) \,m\,,
\end{equation}
if
\begin{equation} \label{cond0}
|\omega|  >> \kappa\,m^2 \,.
\end{equation}
This implies a parabolic approximation of the drift potential $V(m)=-\int D_1(m) dm$,
whose shape is plotted in Fig.~\ref{fig:pot1}
for different values of $T\simeq T_c$, found from the integration of $D_1$ in Eq.~(\ref{D1D2})
and of the linearized expression(\ref{D1linear}), for comparison.
For $ \omega>0$,
one has a confining quadratic potential, while for $\omega<0$ the
parabolic potential is inverted, with an unstable point at $m=0$.

\subsection{Ornstein-Ulhenbeck approximation}

Now, for linear $D_1$ and constant $D_2$,
the exact solution of Eq.~(\ref{FPE}) reads~\cite{risken}
\begin{equation} \label{approx2}
P(m,t|m_0,0) = \frac{1}{\sqrt{2\pi \sigma^2(t)}}
\,\exp \bigl(
 -\frac{[m-m_0\exp(-\omega \,t)]^2 }{2\sigma^2(t)} \bigr)\,,
\end{equation}
where $\sigma^2(t)= \epsilon[1-\exp(-2\omega \,t)]/\omega$.
This solution applies for $\omega >0$  (Ornstein-Uhlenbeck (OU) process) as well as for $\omega<0$,
and is valid as long as the probability distribution remains strongly picked
so that the inequality (\ref{cond0}) holds for any value of $m$ with non-negligible probability.

Performing the average  with Eq.~(\ref{approx2}) gives
\begin{equation} \label{m1a}
\langle m \rangle = m_0 \exp(- \omega \,t) \,.
\end{equation}
Therefore, for $\omega>(<)0$, that is $T/T_c>(<)\,1-1/N$,
the average magnetization decays (grows) exponentially,
with characteristic time $|\omega|^{-1}$. Then, for time scales $t \ll |\omega|^{-1}$, it remains
$\langle m \rangle \sim m_0$.  Since in the large $N$ limit  $\omega \sim\lambda$,
 then  the magnetization scales as $\langle m \rangle = m_0 \,
F(\lambda\, t)$. This is consistent with Eq.~(\ref{eq:short-time-m-scaling}),
provided that $\theta=0$ and   $\nu z=1$, in agreement with the mean-field exponents
$\nu=1/2$ and $z=2$. The same exponents are displayed by the Gaussian model~\cite{JaScSc1989}.
For higher-order moments $m^{(n)} \equiv \left<(m-\left<m\right>)^n \right>$
with even $n\ge 2$,  one has
\begin{equation} \label{mna}
m^{(n)} = \frac{\Gamma(\frac{n+1}{2})}{\sqrt{\pi}} [2\epsilon\,
\omega^{-1} (1-\exp[-2\omega \,t])]^\frac{n}{2} \,.
\end{equation}
Then, for short times $t<<1/|\omega|$,
\begin{equation} \label{scalingmn}
m^{(n)} \sim [\epsilon \,t]^{n/2} \,.
\end{equation}
Hence, $m^{(2)} \sim t/N$, consistently with Eq.~(\ref{eq:scaling-second-moment})
($\beta=1/2$),  provided that we choose $d=4$, the upper critical dimension.

The characteristic time scale for STD behavior is then $t \ll \tau_{STD}$ with
\begin{equation}\label{tauSTD}
    \tau_{STD} \approx \frac{1}{|\lambda+\epsilon|}= \frac{N}{|1+N\,\lambda|} \,.
\end{equation}
If $|\lambda\, N| \gg 1$ we have $\tau_{STD}\sim 1/|\lambda| \ll N$,
while for $|\lambda\, N| \ll 1$ we have $\tau_{STD}\sim N$.

\begin{figure}[h!]
\centering
\includegraphics*[scale=0.4]{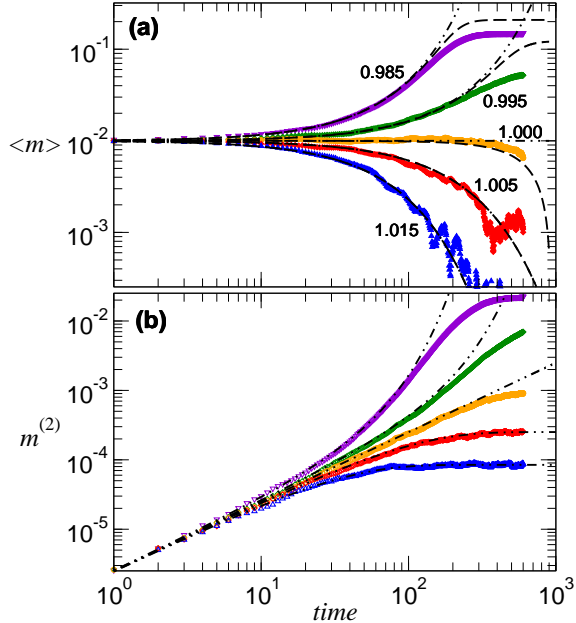}
\caption{(Color online) First and second moments of the order parameter as a function of time $t$,
for $m_0=0.01$ and different values of $T\simeq T_c=1$.
(a) Magnetization: black dot-dashed lines correspond to Eq.~(\ref{m1a}) and black dashed
ones to Eq.~(\ref{m-determ}).
(b) Second moment: black dot-dashed lines correspond to Eq.~(\ref{mna}).
Numerical simulations using Glauber dynamics
were performed for $N= 8\times 10^5$ (color symbols).
}
\label{fig:crit}
\end{figure}

Fig.~\ref{fig:crit} displays the comparison between numerical simulations and
the approximate OU solutions Eqs.~(\ref{m1a})-(\ref{mna}),
for $N= 8\times 10^5$, $m_0=0.01$ and different values of $T\simeq T_c$, such that
$|\lambda\, N| \gg 1$. The OU approximation gives an excellent agreement for time scales up
to $t \sim \tau_{STD}$ ($\tau_{STD}\sim 100$  for the present parameter values).
Averages were taken over 1000 independent MC runs.
The main differences between the theoretical and numerical results appear for $T < T_c$
and $t > \tau_{STD}$, where finite-size effects shift the equilibrium value of both the
average magnetization and its variance.

Fig.~\ref{fig:crit} also shows the performance of Eq.~(\ref{m-determ}), which reproduces
the simulation results for longer times than Eq.~(\ref{m1a}), predicting the transient
steady state. The lower saturation level observed
in simulation outcomes for $T<T_c$ is due to the presence of
fluctuations that drive some trajectories  to the equilibrium state with negative
magnetization, while the deterministic equation rules the stabilization at the level of the local minimum.
Also notice that this discrepancy decreases as $T$ departs from the
critical value because of the consequent increase of the potential barrier height,
which makes such events less probable.
For $T>T_c$, the system evolves quickly towards the vicinity of
the equilibrium state  and the saturation level of the second moment is very close to
the value given by the (bimodal) steady state distribution $P(m)\propto \exp(-V(m)/\epsilon)$.
In any case, finite-size higher order corrections can be neglected
as far as the STD behavior is concerned.

\subsection{Quartic approximation of the drift potential}

When (\ref{cond0}) does not apply, one can not discard the cubic
contribution to $D_1$.  For such case we show in Appendix \ref{appenA} that
the inclusion of  the cubic correction
in the drift coefficient Eq.~(\ref{D1D2_crit}) leads for $\epsilon\ll 1$ to
\begin{equation}\label{m-determ}
   \left< m \right> = \,
 \frac{m_0 {\rm e}^{-\omega t}}{\sqrt{1+ m_0^2 \kappa(1 -  {\rm e}^{-2\omega t})/\omega    }} \,.
\end{equation}
This solution is exact in the thermodynamic limit $\epsilon\to 0$,
as can be verified by direct integration
of the deterministic version of Eq.~(\ref{eqM1})~\cite{binder}, i.e.,
\begin{equation}\label{deterministic}
    \frac{d\langle m\rangle }{dt}=D_1\left( \langle m \rangle \right).
\end{equation}
Notice that the expansion of Eq.~(\ref{m-determ}) up to first order
in $m_0$ reproduces Eq.~(\ref{m1a}).
The case $\omega=0$ ($T=T_c$), can also be drawn from Eq.~(\ref{m-determ})
by taking the limit $\omega \to 0$, yielding
\begin{equation}\label{m-determ0}
   \left< m \right> = \,
 \frac{m_0}  {\sqrt{1+ 2m_0^2 \kappa t    }} \,.
\end{equation}

In Appendix \ref{appenA} we additionally  show that finite-size
corrections do not change the STD scaling of $\left< m\right>$.
For the second moment we obtain
\begin{equation}  \label{m2_spi}
m^{(2)}\equiv \langle m^2 \rangle-\langle m \rangle^2 =
 2\epsilon t\frac{(1+z)(1+2z+2z^2)}{(1+2z)^3} + {\cal O}(\epsilon^2,\epsilon \omega)\,,
\end{equation}
where $z\equiv\kappa m_0^2 t$.
Notice that up to  a typical time scale  $1/(2\kappa m_0^2)$, the approximation
$m^{(2)}\simeq 2\epsilon t$ holds.
For $\kappa m_0^2 t>>1$, a crossover to a second
linear (hence normal diffusive) regime but with a different diffusion constant is predicted, namely
$m^{(2)}\simeq \epsilon t/2$, although it typically falls beyond the STD region.

\begin{figure}[ht!]
\centering
\includegraphics[scale=0.4]{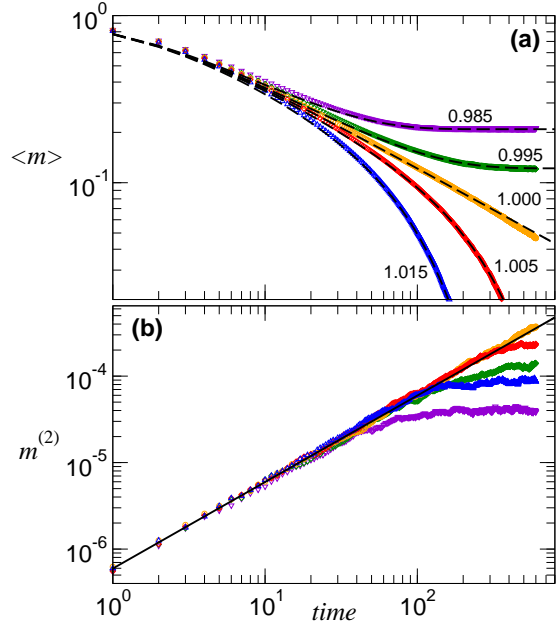}
\caption{(Color online) First and second moments   as a function of time $t$
for $m_0=1$ and different values of $T\simeq T_c=1$.
(a) Magnetization: dashed lines correspond to theoretical results given
by  Eq.~(\ref{m-determ}).
(b) Second moment: the full black line represents $2D_2(m_0=1)t$.
Numerical simulations using Glauber dynamics
were performed for $N= 8\times 10^5$ (color symbols).
}
\label{fig:scaling-m-m01}
\end{figure}

\subsection{Other initial conditions}

To investigate the scaling behavior for other initial
conditions, we analyzed the STD behavior when $m_0=1$.
As can be seen in the inset of Fig.~\ref{fig:pot1}, the cubic approximation
still holds close to $m=1$. Hence, the thermodynamic-limit expression (\ref{m-determ}) is expected
to apply too, as verified in Fig.~\ref{fig:scaling-m-m01}a.
In comparison with the initial condition of Fig.~\ref{fig:crit}, here trajectories get
more trapped around the positive minimum, hence the agreement with deterministic Eq.~(\ref{m-determ}) is
still better.
For finite systems, the intensity of the fluctuations is state dependent
following Eq.~(\ref{D1D2}).
Therefore, the finite-size corrections derived by assuming $D_2\simeq \epsilon$ do not hold.
However, for very short times  one still expects $m^{(2)}\sim 2 D_2(m_0)t$, according to
Eq.~(\ref{M1M21storder}), as in fact verified in numerical
simulations illustrated in Fig.~\ref{fig:scaling-m-m01}.
From Eq.~(\ref{m-determ}) we have that $m(t)\sim t^{-1/2} (1-\lambda t)$ for $t \ll 1/|\lambda|$,
in agreement with Eq.~(\ref{eq:short-time-m-ordered}).
The excellent accord between
Eq.~(\ref{m-determ}) and numerical simulation outcomes displayed in Fig.~\ref{fig:scaling-m-m01}
when $|\lambda\, N| \gg 1$ confirms our previous assumptions. Numerical simulations for other
values of $N$ also verify the above scaling.
For $T > T_c$ the equilibrium (final steady state)
values of both mean and variance are quickly approached as in Fig.\ref{fig:crit}.
However, when $T< T_c$, we see from Fig.\ref{fig:scaling-m-m01}b that all the curves lie below
the critical curve, at variance with the behavior observed when $m_0 \ll 1$
(compare with Fig.\ref{fig:crit}b).
This is because when $m_0 =1$ almost all the trajectories get trapped in the positive minimum.
Thus, the variance stabilizes in a value corresponding to the fluctuations in a single potential
minimum. At long enough times, both minima
in a finite size system get equally populated and therefore the equilibrium value of $m^{(2)}$
will be higher. However, the time scales needed to observe this effect fall outside the STD regime.
On the contrary, when $m_0 \ll 1$, a relatively large number of trajectories cross the barrier
between minima and $m^{(2)}$ approaches the equilibrium value (which is larger than the steady one),
even at very short times, as can be verified by comparing the numerical plateaux
in Fig.\ref{fig:crit}b with the equilibrium value
\begin{equation}
    m^{(2)}_{eq} = \frac{\int_{-1}^1 m^2\, e^{\frac{-V(m)}{\epsilon}}\, dm}{\int_{-1}^1  e^{\frac{-V(m)}{\epsilon}}\, dm} \,.
\end{equation}

%%%%%%%%%%%%%%%%%%%%%%%%%%%%%%%%%%%%%%%
\section{STD near the spinodal}

When $T<T_c$ the model has a line of first-order transitions at $H=0$
and metastable stationary solutions for a range of values of $H$.
Without loss of generality we will restrict hereafter to the metastable solutions with
positive magnetization, that is, those analytic continuations of the equilibrium
magnetization from positive to negative values of $H$. Defining $h\equiv \beta H$,
the metastable state exists as long as $h > h_{SP}$, where the spinodal field is given by
\begin{eqnarray} \nonumber
h_{SP}&=& - \beta J m_{SP}  + \frac{1}{2} \ln \frac{1+m_{SP}}{1-m_{SP}}\\ \nonumber
m_{SP}&=&\sqrt{1-\frac{1}{\beta J}} \,.
\end{eqnarray}
where $m_{SP}$ is the magnetization at the spinodal point~\cite{LoFeCaGr2009}.

Suppose now that we start the system evolution from the completely ordered state $m_0=1$
with $T< T_c$ and $h > h_{SP}$ and let us define $\Delta m\equiv m-m_{SP}$ and
$\Delta h\equiv h-h_{SP}$. Considering $\Delta m$ as an order parameter,
numerical simulations using Metropolis dynamics~\cite{LoFeCaGr2009} showed that
close enough to the spinodal point ($|\Delta h| \ll 1$) its moments obey the
scaling form (\ref{scaling2}) with $\tau= \Delta h/h_{SP}$. For temperatures
far enough from $T_c$ the spinodal magnetization $m_{SP}$ is close to one and we
can expand $D_1$ and $D_2$ in powers of $\Delta h$ and $\Delta m$. Moreover,
close to the spinodal we can neglect~\cite{mori} the finite-size correction of
$D_1$. Then,  from  Eqs.~(\ref{D1D2})
one has at first order in $\Delta h$ and second order in $\Delta m$:
\begin{eqnarray} \nonumber
D_1(m) &\simeq & \frac{\Delta h}{\beta J}
\;-2 m_{SP} \Delta m \Delta h \;-\beta J  m_{SP} (\Delta m)^2  \,,
\\  \nonumber
D_2(m) &\simeq & \epsilon\left( \frac{1}{\beta J} -2m_{SP}\Delta m
+(\beta J-2)(\Delta m)^2 \right.
\\  \label{D1D2_spi}
&& \left. -\frac{m_{SP}}{\beta J}\Delta h
+(2-\frac{3}{\beta J})\Delta m \Delta h
\right)\,.
\end{eqnarray}

\begin{figure}
\centering
\includegraphics*[bb=40 370 520 730, width=0.6\textwidth]{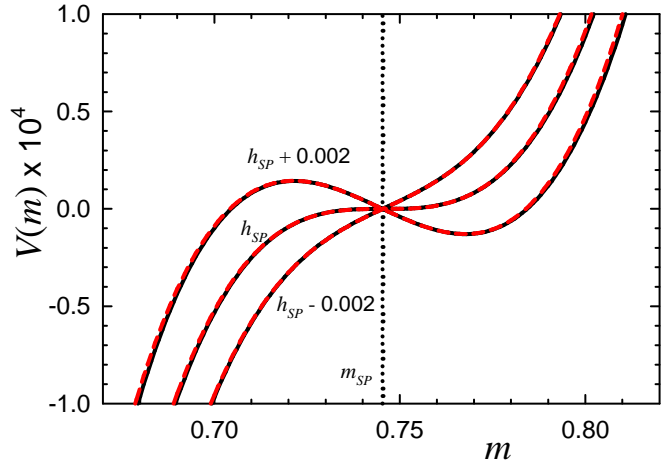}
\caption{ (Color online) Potential $V(m)$, for different values of $h$ in
the vicinity of $h_{SP}$, indicated in the figure for $T=4/9$.
It was computed  (unless an arbitrary additive constant) from
the  integration of $D_1$ in Eq.~(\ref{D1D2}) (red full lines) and (\ref{D1D2_spi})
(black dashed lines).
}
\label{fig:pot2}
\end{figure}

In Fig.~\ref{fig:pot2} we plot the shape of  $V(m)$
for different values of $h$ in the vicinity of $h_{SP}$, obtained both
from integration of $D_1$ in Eq.~(\ref{D1D2}) and of the approximate quadratic polynomial
(\ref{D1D2_spi}), for comparison.

The moments of $\Delta m$ can be calculated by means of  Eq.~(\ref{adjoint}), namely
\begin{equation}  \label{mn_adj}
\langle (\Delta m)^n \rangle  = \sum_{k\ge 0} [D_1\partial_x+D_2\partial_{xx}]^k x^n\,t^k/k! \,,
\end{equation}
where we have defined $x \equiv m_0-m_{SP}$.

For $n=1$ we can neglect in a first approximation the diffusion term, that is,
at least for short times we can disregard finite-size effects.
Then, from Eq.~(\ref{mn_adj}) using
$D_1=-A(x^2+2A\alpha x-\alpha)$,
with $\alpha=  \Delta h/(\beta J \,A)$ and $A \equiv \beta J m_{SP}$
one has (see Appendix \ref{appenB})
\begin{equation}
    \langle \Delta m \rangle  =\sqrt{\gamma}\,\frac{u+\,\tanh(\sqrt{\gamma}At)}
    {1+u\,\tanh(\sqrt{\gamma}At)} -A\alpha\label{m1_dh} \,,
\end{equation}
where $u=(x+A\alpha)/\sqrt{\gamma}$ and $\gamma=\alpha+A^2\alpha^2$.
For $\alpha<0$ (hence $\gamma<0$), Eq.~(\ref{m1_dh}) becomes
\begin{equation}
    \langle \Delta m \rangle  =\sqrt{|\gamma|}\,\frac{u-\,\tan(\sqrt{|\gamma|}At)}
    {1+u\,\tan(\sqrt{|\gamma|}At)} -A\alpha  \label{m1_dh2} \,.
\end{equation}

\begin{figure}[t!]
\centering
\includegraphics*[scale=0.4]{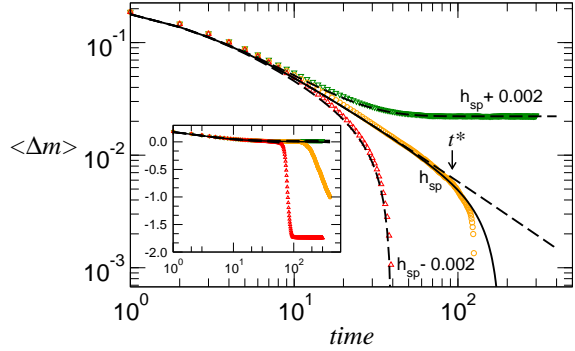}
\caption{(Color online)
Mean magnetization as a function of time $t$, for $m_0=1$,
$T =4/9$ and different values of $h$.
Black dashed lines correspond to the prediction given
by Eqs.~(\ref{m1_dh})-(\ref{m1_spi}). The full black line corresponds
to Eq.~(\ref{m1_epsfull}).
The inset is the same plot in linear-log scales.
Numerical simulations were performed for $N= 8\times 10^5$  (color symbols).
}
\label{fig:spi}
\end{figure}

Alternatively, Eqs.~(\ref{m1_dh})-(\ref{m1_dh2}) can be obtained
by integrating Eq.~(\ref{deterministic}),
and  are  in good agreement with numerical simulations,
as illustrated in Fig.~(\ref{fig:spi}). One observes the following asymptotic behaviors:

(i) For $|h|<|h_{SP}|$ ($\alpha>0$), a constant level is reached. In fact, since
the potential presents a local minimum, the plateau  occurs at a level associated to that minimum.
This is in accord  with numerical simulations (Fig.~\ref{fig:spi}), notice that
 the local minimum of the potential is at $m \simeq 0.768$, then  $\Delta m =m-m_{SP} \simeq 0.023$,
 in agreement with the observed level.

(ii) For $|h|>|h_{SP}|$: ($\alpha<0$), Eq.~(\ref{m1_dh2}) yields a rapid decay towards zero
attained at finite $t$. This is because the potential is tilted towards
the absolute minimum (without local minimum).

%%%%%%%%%%%%%%%%%%%%%%%%%%%%%%%%%%%%%%%%%%%%%%%%%%%%%%%%%%%%%%%%%%%%%%%%%%%%%%%%%%%%%%%%%%%%%%%%%%

In the limit $\alpha \to 0$, from Eq.~(\ref{m1_dh}) it follows
\begin{equation} \label{m1_spi}
\langle \Delta m \rangle  =  \,  \frac{x}{1+Axt}    \,.
\end{equation}
Hence, at the spinodal point one has  $\langle \Delta m(t) \rangle \sim t^{-1}$
for $t\gg 1/Ax$, consistently with Eq.~(\ref{eq:short-time-m-ordered}) with $\beta=1/2$ and $\nu z =1/2$,
in agreement with previous numerical results~\cite{LoFeCaGr2009}. This behavior corresponds
to the relaxation towards the saddle point $m=m_{SP}$.
While in an infinite system such point
is an stationary state, finite-size fluctuations destabilize it, with the subsequent
exponential relaxation towards the equilibrium value $\Delta m \gtrsim -1-h_{SP}$ at
longer times, as depicted in Fig.~\ref{fig:spi}.
Finite-size corrections to Eq.~(\ref{m1_spi}), that we compute for $\Delta h=0$,
can be obtained by including the diffusion term in Eq.~(\ref{mn_adj}).
When  $\Delta h=0$, following Eq.~(\ref{D1D2_spi}), we have
 $D_2(x)\simeq \epsilon(ax^2+bx+c)$, with
$a=\beta J-2$, $b= -2 m_{SP}$, $c=1/\beta J$ and  $D_1=-Ax^2$.
In Appendix \ref{appenB} we obtain Eq.~(\ref{m1_epsfull}), furnishing
 $\langle \Delta m \rangle$ corrected at first order in $\epsilon$,  that for $t\gg 1/Ax$ leads to
\begin{equation}\label{m1_eps_large_t}
   \langle \Delta m  \rangle \sim \frac{1}{A t} \left[1- \frac{\epsilon\,c\, A^2}{10}\, t^3
   + {\cal O}(\epsilon t^2,\epsilon^2)\right] \,.
\end{equation}
Hence, finite-size effects will become relevant only when $t \sim t^*$, with
\begin{equation}\label{t*}
    t^*= \left(\frac{10\beta J}{\epsilon A^2}\right)^{1/3}=\left(\frac{10\, N}{-\lambda}\right)^{1/3}\, ,
\end{equation}
in agreement with the scaling proposed in Ref.~\cite{LoFeCaGr2009}:
$t^* \propto N^{z/d_c}$, with $z=2$ and $d_c=6$.

Finally, let us consider the second moment.
In Appendix \ref{appenB} we obtain Eq.~(\ref{m2_epsfull}),
that gives the  $\epsilon$-correction to $\langle (\Delta m)^2\rangle$.
It allows to compute
$\Delta m^{(2)}=\langle (\Delta m)^2 \rangle -(\langle \Delta m \rangle )^2$, Eq.~(\ref{m2_eps}),
that at short times $t\ll 1/Ax$ leads to
\begin{equation} \label{Deltam2-spi}
    \Delta m^{(2)} \sim   2 \epsilon(a x^2+bx+c)    \, t
    \simeq   2 D(x)\, t\,,
\end{equation}
in accord with Eq.~(\ref{M1M21storder}).

Meanwhile, for $t \gg 1/Ax$, Eq.~(\ref{m2_epsfull}) behaves as
\begin{equation} \label{m2_eps_large_t}
   \langle (\Delta m)^2  \rangle \sim \frac{1}{(A t)^2}
   \left[1 + \frac{\epsilon\,c\, A^2}{5}\, t^3
   + {\cal O}(\epsilon t^2, \epsilon^2)\right] \,.
\end{equation}
Hence from   Eqs.~(\ref{m1_eps_large_t}) and (\ref{m2_eps_large_t}) one gets
\begin{equation} \label{Deltam2-spi-eps}
\Delta m^{(2)}
\sim \frac{2\epsilon c t}{5} \,.
\end{equation}

\begin{figure}[h!]
\centering
\includegraphics*[scale=0.4]{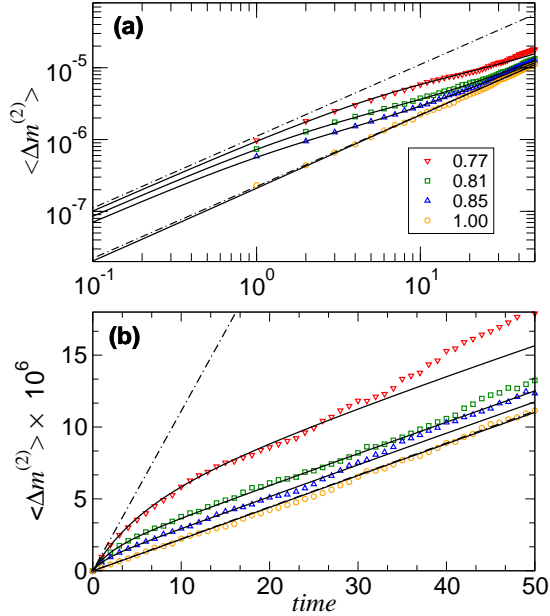}
\caption{(Color online)
Second moment of the order parameter as a function of time $t$,
for $T=4/9$, $h=h_{SP}$ and different values of $m_0$ (with $m_{SP}< m_0<1$).
Panels (a) and (b) display the same data in logarithmic and linear scales, respectively.
Black full lines correspond to the prediction given by  Eq.~(\ref{m2_epsfull}).
Symbols correspond to MC numerical simulations  for $N= 8\times 10^5$. The dash-dotted lines
correspond to  $2 D_2(m_{sp})\, t$ (upper line) and $2/5\,c\,\epsilon\, t$ (lower line).
}
\label{fig:varm0}
\end{figure}

Notice that in this regime,   the prefactor of $t$  given by Eq.~(\ref{Deltam2-spi-eps}) is
generically different from that obtained in the very short-time regime following
Eq.~(\ref{Deltam2-spi}). Fig.~(\ref{fig:varm0}) illustrates this cross-over for different values of $m_0$ and
fixed temperature. The prefactor at small times,
$2D_2(m_0)$ varies with $m_0$ (panel a),
while at intermediate times $1/Ax \ll t < t^*$ the prefactor becomes
$\frac{2}{5} \epsilon c=\frac{2\epsilon}{5\beta J}$ independently of $m_0$,
which is evident in the linear scale (panel b).

In any case the behavior   $\Delta m^{(2)} \sim \epsilon t$ up to $t \sim t^*$ is  consistent  with the STD
scaling hypothesis for the set of mean-field exponents $z \nu =2$, $\beta=1/2$,
$d_c=6$ and in agreement with numerical outcomes~\cite{LoFeCaGr2009}.

%%%%%%%%%%%%%%%%%%%%%%%%%%%%%%%%%%%%%%%
\section{Final comments}

We studied the short-time dynamical behavior of finite-size mean-field models
(infinite-range interactions) with non conserved order parameter dynamics.
By solving the associated Fokker-Plank equation we obtained closed expressions for
the first moments of the order parameter, in the vicinity of both the critical and
spinodal points. This allowed us to confirm  the STD scaling hypothesis in both situations,
as well as to determine
the dynamical ranges of its validity. In particular, we confirmed analytically its validity
 when the system
starts  from an ordered state. Moreover, we found that a diffusion-like scaling behavior of
the second moment appears for any initial value of the order parameter, but the associated
diffusion coefficient presents a crossover between two different values, for short and
intermediate times within the STD regime.

We found in general that the scaling behavior of the first moment is mainly determined by
the shape of the potential $V(m)=-\int D_1(m) dm$ and therefore by the equilibrium
generalized free energy $f(m,T,H)$, which has the same extrema structure as~\cite{binder}
$V(m)$. The scaling behavior of higher moments, on the other hand, has its origin on
the Gaussian nature of finite-size fluctuations close to the singular points.
Although our results were obtained for a particular model, it is worth to stress
that the above facts are characteristic of mean-field systems, since they depend
only on the shape of $V(m)$ and on the proportionality $D_2 \propto 1/N$. This makes
the analysis quite general and independent of the particular mean-field model.

\vspace*{2pc}
\noindent
{\bf Acknowledgments:}
The authors would like to thank T. S. Grigera and E. S. Loscar for sharing with us
their simulation codes for Metropolis dynamics, as well as for useful discussions.
This work was supported by  CNPq and Faperj (Brazil), CONICET,
 Universidad Nacional de C\'ordoba, and ANPCyT/FONCyT (Argentina).

\appendix

\section{Quartic potential approximation near the critical point}
\label{appenA}

To investigate the effect of including the cubic correction  in the drift coefficient
Eq.~(\ref{D1D2_crit}), we evaluate the particular setting of Eq.~(\ref{adjoint})
\begin{equation}\label{adjoint2}
    \langle m^{n} \rangle  =
\sum_{k\ge 0} [(-\omega m_0 -\kappa m_0^3)\partial_{m_0}
+\epsilon \partial_{m_0 m_0}]^k \frac{m_0^n\,t^k}{k!}\,.
\end{equation}
In the limit $|\lambda\, N| \gg 1$, we can neglect in a first
approximation the diffusion term and compute
\begin{equation} \nonumber
\langle m \rangle   \approx
\sum_{k\ge 0} [(-\omega m_0 -\kappa m_0^3)\partial_{m_0}]^k \frac{m_0\,t^k}{k!} \,.
\end{equation}
By iterating the operator $k$ times and  identifying the general form of
the coefficients of $t^k$, with the aid of symbolic manipulation programs,
we obtain
\begin{eqnarray} \nonumber
\langle m \rangle  &\approx& 
 m_0\sum_{k,j\ge0}
\frac{(-\omega t)^k}{k!}
\Bigl(\begin{array}{cc} 2j \\ j  \end{array}\Bigr)
\bigl(-\frac{m_0^2\kappa}{4\omega} \bigr)^j
\sum_{i=0}^j
\Bigl(\begin{array}{cc} j \\ i  \end{array}\Bigr) (-1)^i(2i+1)^k \\ \nonumber
&=&m_0 {\rm e}^{-\omega t}\sum_{j\ge 0}
\Bigl(\begin{array}{cc} 2j \\ j  \end{array}\Bigr)
\bigl( -\frac{m_0^2\kappa }{ 4\omega} (1-{\rm e}^{-2\omega t}) \bigr)^j \\
\,&=& \,
 \frac{m_0 {\rm e}^{-\omega t}}{\sqrt{1+ m_0^2 \kappa(1 -  {\rm e}^{-2\omega t})/\omega    }} \,,
 \label{m-epsilon0}
\end{eqnarray}
that coincides with the exact  deterministic solution (\ref{m-determ}).

Fluctuations can be neglected  as long as $\kappa,\omega \sim O(\epsilon^0)$.
However,  while  $3\kappa$ remains of order one (except for extreme temperatures),
typically $\omega \sim \lambda + \epsilon <<1$.
Then, a finite-size correction can be included by keeping only the terms of order $\epsilon$ and $\omega$
in each coefficient of $t^k$ in Eq.~(\ref{adjoint2}). This procedure yields
the correction term,
\begin{eqnarray} \nonumber
C_1(\epsilon)&=&-\epsilon   \kappa m_0 t^2 \sum_{k\ge 0}
\left(\begin{array}{cc} 2k\\ k  \end{array}\right)
(2k^2+6k+3) (-z/2)^k \\ \nonumber
&=&
-\epsilon \kappa m_0 t^2
\frac{3+4z+2z^2}{(1+2z)^{5/2}} \,,
\end{eqnarray}
where  $z\equiv\kappa m_0^2 t$. Then, it results
\begin{eqnarray} \label{m-epsilon1}
\langle m \rangle &=&
 \frac{m_0 (1- \omega t) }{(1+ 2z)^{1/2}}
+ \frac{  m_0 \omega t\,z}{(1+2z)^{3/2}}
 \\ \nonumber
&&  -\epsilon \kappa m_0 t^2
\frac{3+4z+2z^2}{(1+2z)^{5/2}}
   +{\cal O}(\epsilon^2,\epsilon\omega,\omega^2)  \,.
\end{eqnarray}
Notice that the first two terms in the right-hand side come from the expansion
of the deterministic Eq.~(\ref{m-epsilon0}) up to first order in $\omega$.

In particular, exactly at the critical point we have $\kappa =1/3$ and
$\lambda=0$ (hence $\omega =\epsilon$).
Therefore, as in the case of the OU approximation, one concludes that
the magnetization remains $m \simeq m_0$
up to a characteristic time $\tau_0 \sim 1/\epsilon=N$.

Similarly, for $\langle m^2\rangle$, one obtains the correction
\begin{eqnarray} \nonumber
C_2(\epsilon)&=&
 \epsilon t \sum_{k\ge 0}
(k+1)(k+2)  (-2z)^k =\frac{2\epsilon  t}{(1+2z)^{3}} \,,
\end{eqnarray}
leading to
\begin{equation}
\langle m^2 \rangle =
 \frac{m_0^2}{(1+ 2z) }
- \frac{2\omega z(1+z)}{\kappa(1+2z)^{2}}
+ \frac{2\epsilon  t}{(1+2z)^{3}}
   +{\cal O}(\epsilon^2,\epsilon\omega,\omega^2)  \,.   \label{m-epsilon2}
\end{equation}
Since in the deterministic limit  $\langle m^n\rangle =\langle m\rangle^n$, then
the first two terms in the right-hand side come from the expansion of the squared
Eq.~(\ref{m-epsilon0}) up to first order in $\omega$.

In the computation of the centered second moment, using Eqs.~(\ref{m-epsilon1}) and (\ref{m-epsilon2}),
the purely deterministic terms cancels out to yield  Eq.~(\ref{m2_spi}).

\section{Moment  calculation near the spinodal}
\label{appenB}

If $\epsilon=0$, from Eq.~(\ref{adjoint})  using
$D_1=-A(x^2+2A\alpha x-\alpha)$,
$\alpha=  \Delta h/(\beta J \,A)$ and  $A \equiv \beta J m_{SP}$, the average magnetization is given by
\begin{equation}  \label{m1_adj-epsilon0}
\langle \Delta m \rangle  = \sum_{k\ge 0} [-(x^2+2A\alpha x-\alpha)\, \partial_x]^k x\,(At)^k/k! \,,
\end{equation}
where $x \equiv m_0-m_{SP}$.
Completing squares and making the change of variables $u=(x+A\alpha)/\sqrt{\gamma}$
with  $\gamma= \alpha+A^2\alpha^2$ we obtain
\begin{equation}  \label{m1_adj-epsilon02}
\langle \Delta m \rangle  = \sqrt{\gamma} \sum_{k\ge 0}
[(1-u^2)\, \partial_u]^k u\,(\sqrt{\gamma}At)^k/k! - A \alpha \,.
\end{equation}

Considering the generating function for tangent, with the change of variable $u=\tanh z$,
one has~\cite{tangent}
\begin{eqnarray}
  \sum_{n\ge 0} [ (1-u^2)\partial_u]^n\,u\,\tau^n/n! &=& \sum_{n\ge 0}
  [ \partial_z]^n\tanh z\,\tau^n/n! \nonumber \\
   &=& (u+\tanh \tau)/(1+u\tanh \tau) \nonumber \,,
\end{eqnarray}
from where Eqs.~(\ref{m1_dh})-(\ref{m1_dh2}) follow.

To include finite-size effects we have to consider the complete expression
\begin{equation}  \label{m1full}
\langle \Delta m  \rangle =
\sum_{k\ge 0} [D_1\partial_x+D_2\partial_{xx}]^k \frac{x\,t^k}{k!} \,.
\end{equation}

When $\Delta h =0$, from Eq.~(\ref{D1D2_spi}), we have
 $D_2(x)\simeq \epsilon(ax^2+bx+c)$, with
$a=\beta J-2$, $b= -2 m_{SP}$, $c=1/\beta J$ and  $D_1=-Ax^2$.
The contributions of order $\epsilon$ associated to each coefficient of the
quadratic approximation of $D_2(x)$ are
$$
C_{1a}=-\frac{\epsilon a}{3A} \sum_{k\ge 2}  (k^2 - 1) (-y)^k=
-\frac{ \epsilon a y^2 (3+y)}{3A (1+y)^3} \,,
$$
$$
C_{1b}= \frac{ \epsilon b t}{12 } \sum_{k\ge 2} (k+ 1)(3k-2) (-y)^{k-1} =
-\frac{ \epsilon  b t y  (6+4y +y^2 )}{6 (1+y)^3} \,,
$$
$$
C_{1c}=- \epsilon c A  t^2  \biggl(
1 + \frac{1}{10}\sum_{k\ge 2}  (k + 2)(2k + 1) (-y)^{k - 1}
\biggr)
=-\frac{ \epsilon cAt^2(10+10y+5y^2+y^3)}{10(1+y)^3} \,,
$$
where $y\equiv Axt$.
Summing the $\epsilon$-corrections  $C_{1a}+C_{1b}+C_{1c}$
 together with the deterministic one, given by Eq.~(\ref{m1_spi}),
yields
\begin{eqnarray} \nonumber
&&\langle \Delta m \rangle = \frac{x}{1+y}
-  \left( \frac{c }{10 A x^2}(10+10y+5y^2+y^3) + \right. \\ \label{m1_epsfull}
&& \left. + \frac{b }{6 A x}(6+4y+y^2)  + \frac{a}{3 A}(3+y) \right) \frac{\epsilon \,y^2}{(1+y)^3}\,.
\end{eqnarray}

Likewise, we calculate
\begin{equation}  \label{m2full}
\langle (\Delta m)^2 \rangle =
\sum_{k\ge 0} [D_1\partial_x+D_2\partial_{xx}]^k \frac{x^2\,t^k}{k!} \,.
\end{equation}
In this case, the contributions of order $\epsilon$ are
$$
C_{2a}=-\frac{2ax}{A} \sum_{k\ge 1}
\left(\begin{array}{cc} k+2\\ 3  \end{array}\right)
 (-y)^k= \frac{2ax y }{ A (1+y)^4} \,,
$$
$$
C_{2b}  =  -\frac{b }{12 A} \sum_{k\ge 1}
(k+1)(k+2)(3k+1) (-y)^k  
= \frac{b y (12+6y+4y^2+y^3)}{ 6 A (1+y)^4} \,,
$$
$$
C_{2c}  =   ct
\biggl( 2 +\frac{1}{10} \sum_{k\ge 2}
(k+1)(k+2)(2k+1)
 (-y)^{k-1} \biggr) 
  =  \frac{c  t (10+10y+10y^2 +5y^3+y^4)}{ 5 (1+y)^4} \,.
$$

Summing up the corrections $C_{2a}+C_{2b}+C_{2c}$, together with the deterministic term
(given by the squared Eq.~(\ref{m1_spi})), yields
\begin{eqnarray} \nonumber
 \langle (\Delta m)^2 \rangle &=& \frac{x^2}{(1+y)^2}  
+ \left( \frac{c }{5x}(10+10y+10y^2+5y^3+y^4) \right. \\
&& \left. +\frac{b}{6} (12+6y+4y^2+y^3)  + 2ax \right)\frac{\epsilon\,y}{A(1+y)^4}\,.
\label{m2_epsfull}
\end{eqnarray}

Finally, the second moment is obtained through
$\Delta m^{(2)}=\langle (\Delta m)^2 \rangle -(\langle \Delta m \rangle )^2 $.
The purely deterministic terms cancel out and at first order in $\epsilon$ it
remains
\begin{eqnarray} \nonumber
&&\Delta m^{(2)} = \frac{\epsilon\, y}{30 Ax(1+y)^4} \biggl(
 20 a x^2 (3+3y+y^2) \\ \nonumber
 &+& 15bx(2+y)(2+2y+y^2) + \\  \label{m2_eps}
&+& 12c(5+ 10y +10 y^2 + 5y^3+y^4 ) \biggr) + {\cal O}(\epsilon^2)\,.
\end{eqnarray}

\end{document}